\documentclass[prd,preprint,aps,tightenlines,preprintnumbers,amsmath,amssymb,showpacs,superscriptaddress,longbibliography]{revtex4-1}

\usepackage[T1]{fontenc}
\usepackage[latin9]{inputenc}
\usepackage{float}
\usepackage{bm}
\usepackage{dcolumn}
\usepackage{color}
\usepackage{amsmath}
\usepackage{amssymb}
\usepackage{graphicx} 
\usepackage{hyperref}
\hypersetup{colorlinks,citecolor=blue}
\usepackage[normalem]{ulem}

\graphicspath{{Pictures/}}

\begin{document}

\title{Anomalous anomalies from virtual black holes} 
\author{Joseph Bramante} 
\affiliation{The McDonald Institute and Department of Physics, Engineering Physics, and Astronomy, Queen's University, Kingston, Ontario, K7L 2S8, Canada}
\affiliation{Perimeter Institute for Theoretical Physics, Waterloo, Ontario, N2L 2Y5, Canada}
\author{Elizabeth Gould}
\affiliation{The McDonald Institute and Department of Physics, Engineering Physics, and Astronomy, Queen's University, Kingston, Ontario, K7L 2S8, Canada}
\affiliation{Perimeter Institute for Theoretical Physics, Waterloo, Ontario, N2L 2Y5, Canada}

\date{\today}
\date{\today}

\begin{abstract}
We investigate gravitational UV/IR mixing models that predict a breakdown of low energy effective field theory, from loop-level, non-local gravitational corrections to particle processes. We determine how the choice of IR cutoff in these theories alters predictions for lepton magnetic moments. Using Brookhaven E821 muon magnetic moment data, we exclude models of UV/IR mixing with an IR cutoff set by a spherical volume enclosing the experiment. On the other hand, an IR cutoff defined by the simply-connected spatial volume containing the trajectories of the muons, implies a correction to the muon magnetic moment which may have already been observed.
 \end{abstract}

\maketitle

\section{Introduction}
It is a truth universally acknowledged that physicists, lacking Planck density experiments, do not expect to find quantum gravity in the laboratory. Nevertheless, there are some quantum facets of gravity that arise in regions with sub-Planckian density. 

Indeed, progress is still being made understanding quantum mechanics around black holes, where the energy density is sub-Planckian. Nevertheless, it has become apparent that new quantum effects should arise as a result of gravitational horizons. This spurred the development of black hole radiation, which is too dim to be observed in nature with present technologies \cite{Bekenstein:1973ur,Hawking:1974rv,Page:1976df}. Theoretical analysis of the black hole temperature has led to a set of laws governing black hole thermodynamics, analogous to the known laws of thermodynamics \cite{Bekenstein:1974ax}. The entropy of a black hole's horizon as defined by these laws, has been conjectured as an upper bound on the entropy of any region \cite{Bekenstein:1980jp,Unruh:1982ic,Sorkin:1986mg,Frolov:1993fy,tHooft:1993dmi,Susskind:1994vu,Bousso:1999xy,Marolf:2003wu} (although it is possible to derive an entropy bound without black hole horizons \cite{Casini:2008cr,Marolf:2003sq,Bousso:2004kp}). This implies that a quantum field theory describing a region should somehow incorporate the fact that the entropy of quantum fields in some region cannot exceed the black hole entropy bound.

However, there are reasons to think that quantum field theory's character is altered before the black hole entropy bound is saturated. As identified by Cohen, Kaplan, and Nelson \cite{Cohen:1998zx}, effective field theories describing the dynamics of a region with size $L$ and corresponding infrared cutoff $\Lambda_{IR} \sim \frac{1}{L}$, may break down if the virtual energy density inside that region implies a black hole horizon larger than $L$. This theory also provides a possible resolution of the cosmological constant problem \cite{Cohen:1998zx,Hsu:2004ri,Li:2004rb,Bramante:2019uub}. In effective field theory terms, this ``gravitational UV/IR mixing'' indicates a correspondence between UV and IR cutoffs, which applies going from the UV to the IR, and from the IR to the UV. 

First, let us assume there is an effective theory with ultraviolet cutoff $\Lambda_{UV}$. By definition, this effective field theory must describe energy densities up to $\rho \sim \Lambda_{UV}^4$. But if the effective field theory is applied to a region so large that the energy density $\Lambda_{UV}^4$ implies a black hole, the effective field theory may break down. The size of this region will define an IR cutoff. Approaching from the other direction, let us assume an effective field theory with an infrared cutoff $\Lambda_{IR} \sim \frac{1}{L}$. The Schwarzschild radius of this theory $2 G M $ implies an effective field theory validity bound in terms of the UV cutoff, \mbox{$2 G M = 2 G \rho \frac{4 \pi}{3} L^3  = 2 G \Lambda_{UV}^4 \frac{4 \pi}{3} L^3 < L $}. This restricts the UV cutoff for the theory to
\begin{align}
 \Lambda_{UV}^4 & \lesssim \frac{3}{8 \pi} \frac{M_P^2}{L^2} = \frac{3}{8 \pi} M_P^2 \Lambda_{IR}^2,
 \label{eq:uvir}
\end{align}
where $G = \frac{1}{M_P^2}$, and we will use units with $\hbar = c = 1$. It follows that the UV scale at which this theory breaks down unless black hole states are accounted for is approximately $ \Lambda_{UV} \sim \sqrt{M_P \Lambda_{IR}}$.

It is useful to consider the physics behind the correction arising at $ \Lambda_{UV} \sim \sqrt{M_P \Lambda_{IR}}$. For the moment neglecting field masses, consider the momenta carried by virtual fields in loop-level processes for some experiment with an infrared cutoff $\Lambda_{IR} = \frac{1}{L}$. If the momenta carried in loop-level particle exchanges is greater than $ \Lambda_{UV} \sim \sqrt{M_P \Lambda_{IR}}$, the theory must contain virtual states with densities that exceed black hole densities. Such extremely dense virtual states can be constructed using sufficiently high-order loop processes, by linking together a very large number of virtual fields with momentum $ \Lambda_{UV} \sim \sqrt{M_P \Lambda_{IR}}$. The fact that these states would arise at extremely high order in a perturbation theory does not necessarily mean they can be neglected. In fact, accurately computing virtual gravitational corrections associated with black holes would seem to require a detailed understanding of how quantum fields assemble to form a black hole, which is a topic of active research \cite{Strominger:1996sh,Callan:1996dv,Rovelli:1996dv,Padmanabhan:2009vy,Mathur:2009hf,Papadodimas:2013jku,Harlow:2014yka,Engelhardt:2017aux}.

In the absence of a settled theory for non-local black hole field dynamics, we can attempt an estimate of $\Lambda_{UV}$. Reference \cite{Cohen:1998zx} pointed out that the gravitational ultraviolet/infrared (UV/IR) mixing given by Eq.~\eqref{eq:uvir} may have observable consequences at particle experiments. Corrections to the electron magnetic moment were considered; the electron magnetic moment is arguably the most precisely measured parameter in the Standard Model \cite{Hanneke:2008tm,Parker:2018vye,Aoyama:2012wj,Aoyama:2014sxa}, and is sensitive to corrections from new UV states \cite{Peskin:1995ev}. For the electron anomalous magnetic moment $a_e$, which is related to the electron $g_e-2$ factor by $a_e =\frac{g_e}{2}-1$, the leading order UV and IR contributions from a new state coupled to the electron are
\begin{align}
\delta a_e \simeq \frac{\alpha}{2\pi}\left[\left(\frac{m_e}{\Lambda_{UV}}\right)^2+\left(\frac{1}{m_e L}\right)^2\right],
\label{eq:gcorrect}
\end{align}
where $\alpha$ is the fine structure constant in the Standard Model and $m_e$ is the mass of the electron. In \cite{Cohen:1998zx}, the gravitational UV/IR relation $L = \frac{\Lambda_{UV}^2}{M_P} $ was substituted into the second term in Eq.~\eqref{eq:gcorrect}, and $\Lambda_{UV}$ was varied to obtain the minimum correction to $a_e$,
\begin{align}
\delta a_{e} \Big|_{min} = \frac{\alpha}{2\pi} \left(\frac{m_e}{M_P} \right)^{2/3} \simeq 0.14 \times 10^{-17},
\label{eq:min}
\end{align}
which is five orders of magnitude smaller than the present uncertainty on the electron $a_e$ measurement, which is $ \pm 0.28 \times 10^{-12} (1 \sigma)$ \cite{Hanneke:2008tm}. It has also been considered in \cite{Strumia:2008qa} whether UV/IR effects could lead to an unexpected running of $\alpha$, when comparing measurements of the electron and muon magnetic moments.

On the other hand, Eq.~\eqref{eq:min} only gives the minimum possible correction from the gravitational UV/IR mixing relation $ \Lambda_{UV} \sim \sqrt{M_P \Lambda_{IR}}$. There are other plausible values for the IR cutoff in the gravitational UV/IR relation. Indeed, one might expect the IR cutoff to be set by an experimental length scale. In this paper we investigate a number of choices for $\Lambda_{IR}$, based on some field theoretic arguments and the geometry of particles contained in precision experiments. Using muon $a_e$ measurements, we will exclude the gravitational UV/IR mixing relation for an infrared cutoff determined by the spherical region circumscribing particle trajectories at precision experiments. On the other hand, for an IR cutoff determined by the simply connected volume containing particle trajectories, we find that the anomalous measurement of the muon anomalous magnetic moment \cite{Bennett:2006fi,Blum:2018mom}, indicate that the effect of virtual black hole states may have already been observed. 

\section{Infrared Cutoffs and Nonlocal Gravitational Corrections}

In the absence of an explicit theory treating virtual black holes, we will begin our treatment with a general gravitational UV/IR mixing relation, 
\begin{align}
2 G M = 2 G \Lambda_{UV}^4 V \simeq R,
\label{eq:schw}
\end{align}
where $R$ is the largest radius required to enclose some region of interest, and $V$ is the volume containing fields of interest. The choice of $R$ on the right hand side of this equation can be motivated by the hoop conjecture \cite{Misner:1974qy,Abrahams:1992ru}, which states that a black hole will form if a ring of size $4 \pi GM$ can be spun around some region of interest with mass $M$. We expect our black hole formation threshold to be similarly defined. There are a number of ways we may define the infrared cutoff for a theory which attempts to include non-local gravitational effects.

\subsection{Spherical Infrared Cutoff}
A simple IR cutoff ansatz is to take $V$ as the spherical volume circumscribing the region within which some measured particles are localized. In this case, $V = \frac{4 \pi}{3} R^3$, as was assumed in Eq.~\eqref{eq:uvir}. The UV cutoff obtained by assuming a spherical volume for our IR cutoff region is
\begin{align}
\Lambda_{UV}^{sph} = \left(\frac{3}{8 \pi} \frac{M_P^2}{R^2} \right)^{1/4}.
\label{eq:uvsph}
\end{align}
We will shortly see that a spherical cutoff can be ruled out using precise measurements of the muon's magnetic moment.

\subsection{Simply Connected Volume Infrared Cutoff}
But perhaps using a spherical volume to set the IR cutoff is too simple. We may expect the virtual black hole effect we are looking for to depend on fluctuations of virtual fields in the space around the particles being measured. In most precision experiments, the particles will be confined to a region (usually a ring or beam), but the virtual states are presumably not confined to that region. However, if a path integral formulation applies to corrections associated with black hole formation thresholds, we should expect these virtual black hole forming contributions to be suppressed for paths spanning regions much larger than the immediate vicinity of the measured particles. This would mean these virtual black hole forming contributions would rapidly decrease away from the path of least action defined by, $e.g.$, a particle's trajectory around a ring. Following this logic, we define a simply connected IR cutoff volume $V_{sc}$ as follows: we take all the space in which the system of interest is confined, as well as all points required to connect any two points in the system by a geodesic, which would simply be a straight line path through space in low gravity. In this case, the IR cutoff determined by a simply connected volume gives a UV cutoff
\begin{align}
\Lambda_{UV}^{sc} = \left(\frac{1}{2} \frac{M_P^2 R}{V_{sc}} \right)^{1/4},
\label{eq:uvsc}
\end{align}
where the exact form of $V_{sc}$ will depend on the geometry of the system being analyzed. It is important to note that while this choice of the IR cutoff volume can be non-spherical, it is still a spherical region, as indicated by Eq.~\eqref{eq:schw}, within which virtual fields are restricted by the UV/IR relation.

\section{Implications for Electron and Muon Magnetic Moment Measurements}

We first consider the Penning trap measurement of the electron's anomalous magnetic dipole moment \cite{Hanneke:2008tm}. This experiment consisted of an electron gyrating at $f \approx 150~{\rm Ghz}$ around a $|\vec B| \approx 5.3 ~{\rm Tesla}$ magnetic field, with boost $\gamma_e \approx 3.07$, held in position by an electric quadrupole potential \cite{Hanneke:2010au}. The anomaly $a_e$ was measured using transitions between different electron spin and energy levels. The gyration radius of the trapped electron was $R_e = \frac{v}{ \omega} = {\gamma_e m_e v}{  q B} \approx 0.03 ~{\rm cm}$, where $v,\omega, q$ are the velocity, angular frequency, and charge of the gyrating electron. Using $R_e$, we find that $\Lambda_{UV}^{sph} \simeq 1700~{\rm GeV}$, which predicts a correction to the electron magnetic moment at this experiment
\begin{align}
\delta a_e^{sph} = \frac{\alpha}{2 \pi} \left( \frac{m_e}{\Lambda_{UV}^{sph}} \right)^2 \simeq 10^{-16},
\end{align}
which is more than four orders of magnitude smaller than the present uncertainty on the measurement of $a_e$ \cite{Hanneke:2008tm,Parker:2018vye}. The gravitational UV/IR correction from $\Lambda_{UV}^{sc}$ using a simply connected geometry would be even smaller.

Next we consider the muon $g-2$ experiment E821 conducted at Brookhaven \cite{Bennett:2006fi}. Here the magnetic moment of muons was determined by measuring the cyclotron and spin precession frequencies of the muons. The muons' spin direction and corresponding precession frequency can be extracted using the electron decay direction in $\mu \rightarrow e +\bar \nu_e + \nu_\mu $, which is correlated with the muon spin direction. The muons were entrained in a ring with radius $R_\mu = 711 ~{\rm cm}$, from which we find $\Lambda_{UV}^{sph} \simeq 11~{\rm GeV}$. This predicts a gravitational UV/IR correction to the muons contained in this ring of order
\begin{align}
\delta a_\mu^{sph} = \frac{\alpha}{2 \pi} \left( \frac{m_\mu}{\Lambda_{UV}^{sph}} \right)^2 \simeq 1.1 \times 10^{-7},
\end{align}
where $m_\mu$ is the muon mass. Comparing this to how closely the present measurement of the muon dipole magnetic moment matches Standard Model predictions \cite{Blum:2018mom},
\begin{align}
a_\mu^{EXP}-a_\mu^{SM} = 2.74 \pm 0.73 \times 10^{-9},
\end{align}
we can safely exclude a spherically-defined IR cutoff for gravitational UV/IR mixing.  

Before continuing, a few comments are in order about the size of the UV cutoff we have derived for a spherically-defined gravitational UV/IR mixing relation at the Brookhaven E821 experiment, $\Lambda_{UV}^{sph} \simeq 11~{\rm GeV}$. One might think that new physics at such a low UV scale could be excluded, simply on the grounds that other particle physics experiments like the Large Electron Positron Collider and the Large Hadron Collider, are sensitive even to weakly coupled particle dynamics up to UV cutoffs $\Lambda_{UV} \sim 1000 ~{\rm GeV}$ \cite{Brehmer:2017lrt,Ellis:2018gqa,Bramante:2014hua}. However, it is important to recognize that at these colliders, the relevant experimental volume is either the interaction region of the colliding particles or the region within which a single collision has been ``vertexed.'' This small volume results in a much larger UV cutoff at these experiments. While a detailed analysis of a particular observable would have to be carried out to determine the sensitivity of a high energy collider to gravitational UV/IR mixing, the relevant collision regions are smaller than $100 ~{\rm \mu m}$ at these experiments \cite{Herr:941318,Chatrchyan:2014fea}, which corresponds to a spherically-defined gravitational UV/IR relation cutoff of $\Lambda_{UV}^{collider} \sim 3000 ~{\rm GeV}$. Therefore, it appears gravitational UV/IR mixing may lie beyond the sensitivity of present collider data. 

Next we consider an IR cutoff for the Brookhaven muon $g-2$ experiment, set by using the simply connected volume defined by the muons' trajectories as they circulate in the experiment. Considering the muon Lorentz factor $\gamma_\mu = 29.3$ and a lifetime of $2.2 \times 10^{-6} s$, it is reasonable to assume that each muon circulates hundreds of times around the ring before decaying. The density of material surrounding the muons is low enough, that we assume there is no sizable correction from gravitational curvature to the simply connected volume as defined above. The muons were kept within a few cm of their mean radius, $R_\mu=711.2$ cm. For example, it was reported during the R99 period of data collection, that the vertical variation of the muons was $h \approx 1.55$ cm. We therefore approximate the simply connected volume containing the muon trajectories as
\begin{align}
V_{sc}^{\mu} = \pi R^2 h, 
\end{align}
which corresponds to a UV cutoff
\begin{align}
\Lambda_{UV}^{sc} \simeq \left(\frac{1}{2 \pi} \frac{M_P^2 }{R_\mu h} \right)^{1/4} \simeq 60~{\rm GeV}~ \left(\frac{1~{\rm cm}}{h} \right)^{1/4} \left(\frac{710~{\rm cm}}{R_\mu} \right)^{1/4},
\end{align}
and predicts a muon magnetic moment anomaly correction
\begin{align}
\delta a_\mu^{sc} = \frac{\alpha}{2 \pi} \left( \frac{m_\mu}{\Lambda_{UV}^{sc}} \right)^2 \simeq 3.5 \times 10^{-9} ~\left(\frac{h}{1~{\rm cm}} \right)^{1/2}\left(\frac{R_\mu}{710~{\rm cm}} \right)^{1/2}.
\label{eq:deltasc}
\end{align}
It is interesting that the size of this correction is close to the discrepancy that has been observed at the E821 experiment. One clear way to test the simply-connected gravitational UV/IR mixing model, would be to adjust the electric quadrupole confinement of the muons, so that their trajectories enclose a larger or smaller volume. The effect should increase or decrease as indicated by Eq.~\eqref{eq:deltasc}.

\section{Conclusions}\label{conclude}

We have investigated some observable consequences for gravitational UV/IR mixing, which is a proposed correspondence between a quantum field theory's UV and IR cutoffs, as determined by the energy density at which black holes would form in a region described by the theory. We have found that Brookhaven E821 muon g-2 measurements can exclude the gravitational UV/IR mixing correspondence, for an IR cutoff defined using a spherical volume circumscribing the E821 muon ring.

On the other hand, a UV/IR mixing correspondence determined by using the simply connected region surrounding the volume traversed by E821 muons, has indicated a correction to the muon magnetic moment that is close in size to the presently observed $3.7 \sigma$ deviation from Standard Model predictions \cite{Blum:2018mom}. Because this UV/IR mixing effect would arise from nonlocal dynamics determined by the volume containing the muons, it can be excluded or validated by analyzing muons with larger and smaller trajectories. 

There is a great deal remaining to be explored for gravitational UV/IR mixing. The infrared cutoff volumes defined in this work were applied to experiments conducted in backgrounds with minimal gravitational curvature. Gravitational UV/IR mixing could be explored in regions with higher density, for example the interior of a neutron star. A covariant definition of the ``energy density'' bound explored in this work, might be derived similar to a covariant entropy bound \cite{Bousso:1999xy}. Whether there is a relationship between gravitational UV/IR mixing and UV/IR mixing in noncommutative field theories can be determined \cite{Minwalla:1999px,Gomis:2000zz,Carroll:2001ws,Matusis:2000jf,Craig:2019zbn}, along with the bearing this would have on lepton magnetic moment measurements.

In conclusion, it is intriguing that precision particle experiments have already begun testing theories predicting nonlocal mixing between the UV and IR cutoff of a quantum field theory, as determined by the threshold for black hole formation.

\section*{Acknowledgements}
We acknowledge the support of the Natural Sciences and Engineering Research Council of Canada. Research at Perimeter Institute is supported by the Government of Canada through Industry Canada and by the Province of Ontario through the Ministry of Economic Development \& Innovation.

\bibliography{guvir}

\end{document}